\documentclass[a4paper,12pt]{article}
\usepackage[T1]{fontenc}
\usepackage[letterpaper,tmargin=2.5cm,bmargin=2.5cm,rmargin=2cm,lmargin=2cm]{geometry}
\usepackage{amssymb,amsmath}
\usepackage{graphicx}
\usepackage{fancyhdr}
\usepackage{fancybox}
\usepackage{shadow}
\usepackage{empheq}
\usepackage{titlesec}
\usepackage{titletoc}
\usepackage{soul, color}
\usepackage{float}
\usepackage{shorttoc}
\usepackage{xcolor}
\usepackage{fancybox}
\usepackage{natbib}
\usepackage{array}
\usepackage{tabularx}
\usepackage{booktabs}
\usepackage{rotating}
\usepackage{bigstrut}
\usepackage{natbib}
\usepackage{aeguill}
\everymath{\displaystyle}
\usepackage{subfigure}

\newtheorem{prop}{Proposition}

\usepackage{changepage}

\title{A Note on the Topology of the First Stage of 2SLS with Many Instruments}
\author{Guy Tchuente\thanks{School of Economics and MaGHiC, Email: guytchuente@gmail.com. The author thanks anonymous referees and the Associate Editor for very thoughtful comments that have improved the current state of the paper. This paper is currently ongoing an in-depth revision for Reject and Resubmit at the JBES, please e-mail the author for comments and suggestions.}\\
University of Kent}
\date{ July 2021}
\begin{document}
\maketitle
\abstract{Finite sample properties of estimators are usually understood or approximated using asymptotic theories. Two main asymptotic constructions have been used to characterize the presence of many instruments. The first assumes that the number of instruments increases with the sample size. I demonstrate that in this case, one of the key assumptions used in the asymptotic construction may imply that the number of ``effective" instruments should be finite, resulting in an internal contradiction. The second asymptotic representation considers that the number of instrumental variables (IVs) may be finite, infinite or even a continuum. The number does not change with the sample size. In this scenario, the regularized estimator obtained depends on the topology imposed on the set of instruments as well as on a regularization parameter. These restrictions may induce a bias or restrict the set of admissible instruments. However, the assumptions are internally coherent. The limitations of many IVs asymptotic  assumptions provide support for finite sample distributional studies to better understand the behavior of many IV estimators. 
 }

\section{Introduction}
This paper discusses implications of using many IVs (IVs) in an asymptotic setting.  I illustrate the many IVs asymptotic framework with the following simple linear model:
\begin{eqnarray}
  \underset{N\times 1}{y} &=& \underset{N\times G}{X}\underset{G\times 1}{\delta}+\underset{N\times 1}{u},  \\
 X & = & \underset{N\times G}{f(Z)}+\underset{N\times G}{V}.
\end{eqnarray}

Empirical researchers are usually interested in the effect of a set of variables, $X$, on an outcome, $y$. I assume endogeneity of $X$ with respect to $\delta$ (i.e. $X$ is correlated with the error term $u$). To solve the endogeneity problem, some IVs ($Z$) are identified. Having many instruments may be desirable as it could increase the fit of Equation (2).  However, the precision improvement comes at the expense of bias in the estimation of $\delta$ using the IVs (see \cite{Bekker94}). The behavior of the estimator in a finite sample is usually approximated using asymptotic theories. An important question in theoretical econometrics is how to formalize the presence of many instruments in an asymptotic construction. In the many IVs literature, there are two main branches, differentiated by the characterization of the IVs set.

The first and most important part of the theoretical econometrics literature assumes that the number of instruments, represented by the number of columns of $Z$ ($K$), grows with sample size $N$. The asymptotic construction is done \`{a} la \cite{Bekker94}, i.e. with $\frac{K}{N}\longrightarrow \kappa$ or with a moderate number of instruments $\frac{K^d}{N}\longrightarrow 0 $, where $d=2$ or $3$. The value of $d$ depends on whether asymptotic convergence in probability or in distribution is being investigated; see \cite{andrews2007testing}, \cite{van2010many}, and \cite{anatolyev2011specification}. In all these cases, the number of instruments increases with the sample size.

A second strand of the literature considers the number of instruments as given and unchanging with sample size. It proposes a framework in which estimation can be done with fixed, infinite or a continuum of IVs. Seminal papers are \cite{carrascoandFlorens2000} for generalized method of moments (GMM) with a continuum of moments condition, and \cite{CARRASCO2012} with regularized two-stage least squares (2SLS).

In  this paper, I use  Bekker's asymptotic approximation to refer to the case where the number of IVs grows with the sample size and Carrasco's asymptotic approximation to refer to the case where there could be many, infinite or a continuum of IVs. The aim of this paper is to understand the implications of each approach's  key assumptions.

Using the 2SLS estimators (the regularized and the classical 2SLS), I present an intuitive description of the challenges in using both types of many IVs in an asymptotic environment.  Indeed, the 2SLS estimator is the result of two optimization processes. The first-stage optimization problem is to find $K$ parameters to minimize the sum of the quadratic error in Equation (2). The second stage replaces $X$ with its predicted value from estimates obtained in the first stage.

In Bekker's style of asymptotic approximation, the number of instruments increases with the sample size. This means that the parameter space of the first-stage estimation changes with more observations as the sample becomes closer to the population size. This leads to changes in the first-stage estimation that are not comparable, as the numbers of parameters $K$ are different when the sample size increases. What are the consequences of assuming a growing first stage on the optimization process leading to the IV estimators? I show that certain assumptions proposed in the literature resolve the problem described above by implying implicitly that the number of effective IVs is finite.

In Carrasco's asymptotic approximation, as the sample size approaches the population size, the first-stage optimization problem's dimension stays the same. However, the problem may be infinite-dimensional, in the sense that the object of interest in the first stage belongs to a space with infinite dimensions.  The infinite dimensionality of the problem implies that some assumptions should be imposed on the set of instruments, and regularization methods are needed to solve the problem. \cite{CARRASCO2012} imposes an assumption about the compactness of the IVs covariance operator. Thus, the existence of a regularized estimator depends on the specific topology imposed on the space spanned by the IVs.   Nevertheless, when the instruments are correlated, the assumption is plausible and the set of assumptions are internally coherent. 

I show that the compactness of the covariance operator assumption implies a key assumption of the behavior of instruments in the Bekker's asymptotic approximation. In addition, the compactness assumption  does not hold for orthonormal IVs. In conclusion, the drawbacks associated with these two asymptotic constructions suggests that finite sample distributional investigation may be better for many IVs, consistent with \cite{harding2016finite} and \cite{bun2011comparison}.

\section{Many Instruments Asymptotic Approximation}

This section presents the assumptions commonly used in the many IVs literature. I first discuss the assumptions used in Bekker's asymptotic approximation method, followed by the assumptions in Carrasco's many IVs method. My main focus is the restrictions they imply as well as possible links between the two types of many IVs.

In Bekker's many instruments method, the asymptotic behavior of the estimator of $\delta$ is usually obtained under the following assumptions:

\textbf{Assumption 1:} ${ y_i, X_i, Z_i, u_i, V_i}$, ${i=1,...,N}$ are $iid$. $u_i$ and $V_i$ have a mean of zero and finite fourth moments, and the variance of $(u_i, V_i)$ is non-singular.

 Let us define the following quantities $\sigma_{Vu}=E(V_i'u_i),$ $\sigma_{u}^2=E(u_i^2),$ $\gamma=\sigma_{Vu}/\sigma_{u}^2$ and $\tilde{V}=V-u\gamma'$.

\textbf{Assumption 2:}\\ 
($i$) $E(u_i|\tilde{V},Z_i)=0$,   $E(u_i^2|\tilde{V},Z_i)=\sigma_u^2$ and for some $p>2$, $E(\mid u_i \mid^p |\tilde{V},Z_i)$ is bounded; and\\($ii$) $Z'Z$ is non-singular; $f(Z)=Z\pi$.\\

Assumptions 1 and 2 (i) are common to both Bekker's and Carrasco's approaches. Assumption 2 (ii) has an infinite-dimensional counterpart. 

\textbf{Assumption 3:} \\
($i$) $\frac{K}{N}\longrightarrow \kappa$, with $0\leq \kappa < 1$;\\
($ii$)  $\pi'Z'Z\pi/N \longrightarrow Q$, where $Q$ is a positive definite matrix; and\\ ($iii$) $ max_{i\leq N}\parallel\pi Z'_i\parallel/\sqrt{N} \longrightarrow 0 $.

 Assumption 3 creates the gap between the two strands of the literature on many instruments. The crucial regularity assumption is that  $\pi'Z'Z\pi/N \longrightarrow Q$, where $Q$ is a positive definite matrix.

The 2SLS estimator is often employed in the presence of many instruments. It will be used to show  the importance of these assumptions in the derivation of their asymptotic behavior.
The 2SLS estimator of $\delta$  given by:
\begin{equation}\label{2slsest}
  \hat{\delta}=(X'P_Z X)^{-1}X'P_Zy, 
\end{equation}

where $P_Z=Z(Z'Z)^{-1}Z'$ is the projection matrix on the space spanned by the instrument.

The next subsection investigates the implications of Assumption 3 on the asymptotic behavior of the 2SLS estimator.

\subsection{Implications of Bekker's Asymptotic Approximation Assumptions}
Using Equation (1), I have
$$\hat{\delta}-\delta=(X'P_Z X)^{-1}X'P_Zu= [X'Z(Z'Z)^{-1}Z' X]^{-1}X'Z(Z'Z)^{-1}Z'u.$$
 To prove the consistency of 2SLS with a fixed number of IVs, the law of large numbers implies that $\frac{1}{N}\sum_{i=1}^NZ_i'Z_i \rightarrow^p E(Z_i'Z_i)$, where $E(Z_i'Z_i)$ is finite and nonsingular by assumption. If I assume that $K$ goes to infinity, $E(Z_i'Z_i)$ will become infinite dimensional.  To avoid dealing with this infinite-dimensional object, I use a sequential asymptotic approach. I let the sample size and then $K$ each go to infinity. The sequential asymptotic approximation is applied under Assumption 3. Indeed, \cite{stock2005asymptotic} show, using \cite{phillips1999linear}, that the sequential asymptotic approximation is equivalent to the joint asymptotic approximation (i.e. $K, N \rightarrow \infty$).

 Assumption 3 ($ii$) says that  $\pi'Z'Z\pi/N \longrightarrow Q$, where $Q$ is a positive definite matrix.  This is a  strong assumption when joint asymptotic approximation is considered. For Assumption 3 to be coherent with joint asymptotic approximation, the same value for $Q$ should be obtained regardless of whether the convergence is sequential. 
 
Indeed, in such cases, $\pi'Z'Z\pi/N =\pi'\frac{Z'Z}{N}\pi $  when $N$ and $K$ go to infinity simultaneously, and $\pi$ and  $\frac{Z'Z}{N}$ become infinite dimensional objects. 
 
The law of large numbers implies that $\frac{Z'Z}{N}$ converges to an  infinite dimensional matrix $Q_{\infty}$, the convergence norm needs to be specified.\footnote{As I have infinite dimensions, the topology or the distance used for convergence is important.}

Assuming that the quadratic  product converges to $Q$ regardless of the topology or the norm used can be a problem, because in infinite dimensions all norms are not equivalent. 

Van Hasselt (2010) notes that this assumption implicitly imposes on $Z$; however, he does not discuss the implication of the infinite dimensional nature of the imposed structure and the necessity of discussing the specific norm used. 

For a better understanding of the consequences of different strands assumption I introduce the following definition.

 \textbf{Definition: effective instrument}
Let consider $Z_{K_a}$ the $K^{th}$ instrument vector and, $\pi_{K_a}$ the unobserved coefficient associated to the instrument $Z_{K_a}$ in the linear representation of the reduced form in Equation (2), $|\pi_{K_a}|=O(c)$.  If $\pi_{K_a}=0$, then the instrument $Z_{K_a}$ is irrelevant.   If $c > \frac{1}{\sqrt{N}}$, the instrument is "effective".

\begin{prop}
Under Assumptions 1 to 3, the existence of $Q$ within a joint asymptotic framework implies that the number of effective IVs is finite.
\end{prop}

\textbf{Proof of Proposition 1.}\\
The following proof uses a contradiction argument to show that the existence of $Q$ implies that the number of instruments is finite.
The steps are the following:
\begin{enumerate}
    \item Assume that the number of effective instruments is infinite.
    \item Assume that the joint and sequential convergences have the same limit.
    \item Show that a sequence of matrices indexed by $K$, to obtain 
    Assumption 3, cannot be a Cauchy sequence, and thus cannot be convergent.\footnote{ A sequence ($a_K$), $K>0$ is a Cauchy sequence, in a space with norm $\| \|$, if for all arbitrary small positive real number $\varepsilon$ there exist an integer $K_{\varepsilon}$ such that for any $n,m>K_{\varepsilon}$,  $\| a_n-a_m\|<\varepsilon$. }
\end{enumerate}
Since $\pi$ is supposed to be known for any value of $K$ as it goes to infinity, I can denote $\pi_{\infty}$ as its limit value.
Based on the equivalence between joint and sequential convergence, $Q$ exists if and only if
\begin{eqnarray}\label{equiv}
Q&=& \pi_{\infty}'M_{\infty}\pi_{\infty}\\
&=&plim_{K,N\rightarrow \infty }\pi_{K.}'M_{K}\pi_{K.}
\end{eqnarray}

 where $M_{K}=plim_{N\rightarrow \infty} \frac{Z'Z}{N} $ is a $K\times K$ matrix and $\pi_{K.}=(\pi_1, \pi_2,....,\pi_K)'$ a $K\times G$ matrix.

The quantity $plim_{K,N\rightarrow \infty }\pi_{K.}'M_{K}\pi_{K.}$ is obtained using  a joint asymptotic approximation. It should converge to $\pi_{\infty}'M_{\infty}\pi_{\infty}$, where $\pi_{\infty}$ is an infinite-dimensional vector and $Q_{\infty}$ is an infinite-dimensional matrix.

For all values of $K$, the matrix $\pi_{K.}'M_{K}\pi_{K.}$ belongs to space of $G \times G$ matrices, $\mathbb{M}(G)$. $(\mathbb{M}(G), \|. \|)$,  where $\|. \|$ is the  Frobenius norm, is a complete finite-dimensional space. Thus, all Cauchy sequences are convergent and vice versa.

The discussion below shows that  $\pi_{K.}'M_{K}\pi_{K.}$ cannot be a Cauchy sequence in $K$ unless the number of relevant instruments is fixed. Thus, the convergence of $\pi'\frac{Z'Z}{N}\pi$ can generally only occur with a fixed number of instruments.

\textbf{Step 1:} The number of effective instruments is assumed to be infinite.\\
This implies that for any $K_0>0$, there exists a $K_1>K_0$  such that $\pi_{K_1}\neq 0$.

\textbf{Step 2:} Joint and sequential convergences have the same limit.\\
As a consequence,  $plim_{K\rightarrow \infty }\pi_{K.}'M_{K}\pi_{K.}$ is convergent and has to be a Cauchy sequence.

\textbf{Step 3:} Contradiction: I show that $plim_{K\rightarrow \infty }\pi_{K.}'M_{K}\pi_{K.}$ diverges.\\
  Indeed, I prove that the sequence $\pi_{K.}'M_{K}\pi_{K.}$ is not a Cauchy sequence. 

Consider  any integer $K_L$. If I take $K>K_L$, by assumption, there exists an infinite number of $K_j>K$ such that $\pi_{K_j}\neq 0$ with $j>1$, and for at least one $K_j$, I can show that $\|\pi_{K_j.}'M_{K_j}\pi_{K_j.}-\pi_{K.}'M_{K}\pi_{K.}\|$ does not neighbor  zero.  In other words, I can find an arbitrary small number $\varepsilon_{K_M}$ such that $\|\pi_{K_j.}'M_{K_j}\pi_{K_j.}-\pi_{K.}'M_{K}\pi_{K.}\|>\varepsilon_{K_M}$.

Indeed, let us consider $K_2$ as the minimum of $K_j>K$ such that $|\pi_{K_j}|\geq \frac{1}{\sqrt{N}}$.

Consider the block matrix notation of $M_{K_2}=\left(
                                                                                                                               \begin{array}{cc}
                                                                                                                                 M_{K} & S_{K,K_2} ' \\
                                                                                                                                 S_{K,K_2} & M_{K,K_2} \\
                                                                                                                               \end{array}
                                                                                                                             \right),
$ and  $\pi_{K_2.}=(\pi_{K.}', \pi_{K+1},...,\pi_{K_2})'$.
\begin{eqnarray*}
                                                                                                        \pi_{K_2.}'Q_{K_2}\pi_{K_2.} &=& (\pi_{K.}', \pi_{K+1},...,\pi_{K_2})\left(
                                                                                                                               \begin{array}{cc}
                                                                                                                                 M_{K} & S_{K,K_2} ' \\
                                                                                                                                 S_{K,K_2} & M_{K,K_2} \\
                                                                                                                               \end{array}
                                                                                                                             \right)(\pi_{K.}', \pi_{K+1},...,\pi_{K_2})' \\
                                                                                                                                &=& \pi_{K.}'M_{K}\pi_{K.}+ \pi_{K,K_2}'S_{K,K_2}\pi_{K.}+ ( \pi_{K,K_2}'S_{K,K_2}\pi_{K.})'+\pi_{K,K_2}'M_{K,K_2}\pi_{K,K_2}
        \end{eqnarray*}
with $\pi_{K,K_2}=(\pi_{K+1},...,\pi_{K_2})'$ and  $M_{K,K_2}$ is a square matrix $(K_2-K) \times (K_2-K)$.

I can therefore conclude that   \begin{eqnarray}
                                                                                                               \label{cauch}                                                                         \nonumber              \|\pi_{K_2.}'M_{K_2}\pi_{K_2.}-\pi_{K.}'M_{K}\pi_{K.}\| &=& \|\pi_{K,K_2}'S_{K,K_2}\pi_{K_2.}+ ( \pi_{K,K_2}'S_{K,K_2}\pi_{K_2.})'+\pi_{K,K_2}'M_{K,K_2}\pi_{K,K_2}\| \\
                                                                                                                                                                                                       &=& \|\pi_{K_2.}'\left(
                                                                                                                               \begin{array}{cc}
                                                                                                                                 0 & S_{K,K_2} ' \\
                                                                                                                                 S_{K,K_2} & M_{K,K_2} \\
                                                                                                                               \end{array}
                                                                                                                             \right) \pi_{K_2.}\|.
 \end{eqnarray}
 Let us assume that the IVs are orthogonal and non-constant. Then, $S_{K,K_2}=0$, which implies that  $$\|\pi_{K.}'M_{K}\pi_{K.}-\pi_{K_2.}'M_{K_2}\pi_{K_2.}\| \geq O_p(\frac{1}{N})>0.$$ 
 The above inequality is true because $ M_{K,K_2}$ is a symmetric positive semi-definite matrix and $|\pi_{K_2}|\geq \sqrt{N}$. In this case, $\pi_{K.}'M_{K}\pi_{K.}$ cannot be a Cauchy sequence.

 Similarly, in the case where the new  IVs ($K_2-K$ IVs) are orthogonal to the $K$ first instrument, then $S_{K,K_2}=0$ and thus $$\|\pi_{K.}'M_{K}\pi_{K.}-\pi_{K_2.}'M_{K_2}\pi_{K_2.}\| \geq O_p(\frac{1}{N})>0,$$ leading to a similar conclusion.

 For any set of IVs, I can define  $$A_{K_2}= \left( \begin{array}{cc}
                             0 & S_{K,K_2} ' \\
                             S_{K,K_2} & M_{K,K_2} \\
                             \end{array}
                             \right).$$
 $A_{K_2}$ is a symmetric matrix. Therefore, from the Spectral Theorem, there is a spectral decomposition $A_{K_2}=ODO'$ where $D$ is a diagonal matrix of eigenvalues and $O$ is a matrix of orthonormal eigenvectors. Thus from Equation (\ref{cauch}) and the representation of $A_{K_2}$ above,
$$\|\pi_{K_2.}'M_{K_2}\pi_{K_2.}-\pi_{K.}'M_{K}\pi_{K.}\|=\|\pi_{K_2.}'ODO'\pi_{K_2.}\|$$ which is equal to zero if and only if  $D^{\frac{1}{2}}O'\pi_{K_2.}=0$ (i.e. $\pi_{K_2.}=0$, since $OD^{-\frac{1}{2}}D^{\frac{1}{2}}O'=I$). Also, because $\pi_{K_2}\neq O(\frac{1}{\sqrt{N}})$, $\|\pi_{K_2.}'M_{K_2}\pi_{K_2.}-\pi_{K.}'M_{K}\pi_{K.}\|\geq O_p(\frac{1}{N})>0$. 

Because of the density of the space of real number, showing that $\|\pi_{K_2.}'M_{K_2}\pi_{K_2.}-\pi_{K.}'M_{K}\pi_{K.}\|\geq O_p(\frac{1}{N})>0$ implies that there exist a real number $\varepsilon_{K_L}$ such that $$\|\pi_{K_2.}'M_{K_2}\pi_{K_2.}-\pi_{K.}'M_{K}\pi_{K.}\|\geq O_p(\frac{1}{N})>\varepsilon_{K_L}>0.$$ for all $K$ and $K_2$.

This means that $\pi_{K.}'M_{K}\pi_{K.}$ is not a Cauchy sequence. Indeed, for any given integer $K_L$,  I can find a small real number $\varepsilon_{K_L}$, such that there are $K,K_2>K_L$ with $\|\pi_{K_2.}'M_{K_2}\pi_{K_2.}-\pi_{K.}'M_{K}\pi_{K.}\|>\varepsilon_{K_L}.$

In conclusion, if for any $K_0>0$ there exists a $K_1>K_0$  such that $\pi_{K_1}\geq O_p(\frac{1}{N}) $; then,  $\pi_{K.}'M_{K}\pi_{K.}$ cannot be a Cauchy sequence.  It follows that there needs to be a fixed integer $K_L$ such that for all $K>K_L$, $|\pi_{K}|<O(\frac{1}{N})$.

This ends the proof of Proposition 1.\\

I have shown that Assumption 3 does not hold if joint asymptotic approximation is applied. However, it is a reasonable assumption under sequential asymptotic approximation. The equivalence between the two modes of convergence does not hold in this case because the probability spaces of $Q$ and $\pi_{K.}'M_{K}\pi_{K.}$ are different.\footnote{ $Q$ is the limit in probability of a quantity, and the probability used in that limit calculation is different  from that induced by the random variables used in the calculation of $\pi_{K.}'M_{K}\pi_{K.}$. } Therefore, Lemma 6 of \cite{phillips1999linear} cannot be applied. Indeed, the difference in the probability spaces reflects the fact that the first-stage regression parameter dimension changes as the sample size increases. As the sample size goes to infinity, it becomes difficult to establish the consistency of the estimator of $\pi$ without further assumptions.  Indeed, it can be noted that if $\pi_{K_2}$ go to zero with $N$ the term in Equation (\ref{cauch}) will also go to zero.  This will be possible in the context of \cite{Staigerandstock97} weak instruments where $\pi=\frac{1}{\sqrt{N}}(1,...,1,...)$, or in the situation of \cite{belloni2012sparse} where the first-stage is assumed to be sparse.

Assumption 3 ($ii$) implies that the number of instruments should be fixed (even if large). Intuitively, this result  means that as the sample size increases, there is a point at which a full knowledge of the first-stage equation is achieved and no more IVs are needed. However, in a finite sample the number of instruments can be close to the sample size. This will lead to the many instruments bias problem, which is usually solved by introducing bias correction terms to obtain a bias-corrected 2SLS estimator. Assumption 3, under which many Bekker's type asymptotic approximation of the estimators are constructed, suggests a fixed number of instruments. Therefore, it is Likely  that these asymptotic estimators may not be very good approximation of finite sample behavior of the 2SLS estimator, as pointed out by \cite{harding2016finite}, who  use a finite sample approximation as a solution to the many IVs problem. However, alternative asymptotic estimators could be helpful for inference in large samples. One candidate is Carrasco's asymptotic approximation, which does not assume an increasing number of IVs. The next subsection discusses the assumptions imposed by Carrasco's asymptotic approximation.

\subsection{Implications of the Carrasco's Asymptotic Approximation Assumptions}

In some empirical applications, the structural equation can suggest a form of Equation (1) implying the use of many IVs. Examples of such models include the spatial model and dynamic panel. In these frameworks, the first-stage equation results in an infinite sum. Equation (2) becomes
$$X = \underset{N\times G}{\sum_{j=1}^{\infty} Z(j)\pi_j}+\underset{N\times G}{V}.$$

The present case contains infinite potential IVs, so the 2SLS estimator proposed above cannot be used. Indeed, the number of moment conditions generated by the model is infinite.  In general, if the instruments are assumed to be strictly exogenous, (following \cite{CARRASCO2012}, this means that for $E(u|z)=0$), any function of $z$ can serve as an instrument. For instance, the set of $Z_i(\tau)=exp(i\tau' z_i)$ with $\tau \in \mathbb{R}^b$ and $z_i$ a vector of exogenous variables from  $\mathbb{R}^b$ can be considered as a set of instruments. \cite{CARRASCO2012} proposes a regularized 2SLS estimator for this type of many instruments problem. She uses a unifying framework in which fixed, infinite and a continuum of moment conditions can be examined simultaneously. To account for the high dimensional nature of the problem in the first-stage equation, she proposes four regularization schemes: Tikhonov (ridge), Landweber-Fridman, principal components and spectral cut-off. This section uses the Tikhonov regularization to investigate the implications of the assumptions proposed in this alternative asymptotic approximation framework.

To use regularization methods, the matrix $Z'Z/N$ is replaced by the covariance operator $\mathbf{\Upsilon}$. In particular, for a continuum of moment conditions, the covariance operator is defined as:
\begin{equation}\label{operatL}
\mathbf{\Upsilon}:L^{2}(\mu )\hspace{0.03in}\rightarrow \hspace{0.03in}L^{2}(\mu )
\end{equation}%
\begin{equation*}
(\mathbf{\Upsilon}g)(\tau _{1})=\int E(Z_i(\tau _{1})\overline{Z_i(\tau _{2})})g(\tau _{2})\mu (\tau _{2})d\tau _{2}
\end{equation*}%
where $\overline{Z_i(\tau _{2})}$ denotes the complex conjugate of $
Z_i(\tau _{2})$ and $\pi$ is a probability density function.   Details on the regularized 2SLS estimator are in \cite{CARRASCO2012}, and its value is given by
$$\hat{\delta}_R=(X'P^{\alpha} X)^{-1}X'P^{{\alpha}}y$$
where $P^{\alpha}$ is the infinite-dimensional analogue of the projection matrix on the space spanned by IVs, and $\alpha$ is the regularization parameter. I consider that $f$ function from the space of exogenous variables to $\mathbb{R}^G$. The asymptotic behavior of $\hat{\delta}_R$ is obtained under the following assumptions:\\

\textbf{Assumption 4:} $E[f(z_i)f(z_i)']$ exists and is non-singular, and $f(z_i)$ belongs to the closure of the linear span of  \{$Z_i(\tau)$ with $\tau \in \mathbb{R}$\}.\\

Assumption 4 is the infinite-dimensional counterpart  of Assumption 2 ($ii$). \\

\textbf{Assumption 5:} $\mathbf{\Upsilon}$ is a compact or nuclear operator.\footnote{Let $(X,\big<.,.\big>_X)$ and $(Y,\big<.,.\big>_Y)$ be separable Hilbert spaces. An operator T:$X \rightarrow Y$ is nuclear if it can be represented by $Tx=\sum_{j=1}^{\infty}a_j\big< b_j,x\big>$ for all $x \in X$, where the sequences $\{a_j\} \subset X$ and $\{b_j\} \subset Y$  such that $\sum_{j=1}^{\infty}\|a_j\|_X\|b_j\|_Y < \infty$. }

The new assumption introduced in this case is  that the operator  $\mathbf{\Upsilon}$ is compact (see \cite{CARRASCO2012}) or nuclear (see \cite{carrasco2016efficient}).

\begin{prop}
Assume that Assumptions 1, 2 ($i$)  and 4 hold. Then, Assumption 5 implies Assumption 3 ($ii$). 
\end{prop}
 
\textbf{Proof of Proposition 2.}\\
I explore different numbers of instruments to prove the proposition in each case. The case of finite instruments is trivial: both assumptions 3 ($ii$) and 5 are implied by Assumptions 1, 2, and 4.

In the infinite-dimensional cases, the compactness assumption implies that the spectrum of the operator is discrete.  In other words, the operator can be approximated by a finite-dimensional operator. In practice, this assumption is crucial as it allows for a finite sum to represent the operator. \\

\textbf{($i$) Finite number of instruments}\\

Assumption 4 implies that $f(Z)=Z\pi_0$.  The covariance operator in this case is compact, so Assumptions 3 (ii) and 5 are trivially satisfied.

\textbf{($ii$) Infinite countable number of instruments}\\
I consider that for each individual $i$, $Z_i$ belongs to the space of squared summable sequence $l^2$. Let $x$ and $y$ be two elements of $l^2$. The scalar product $<x,y>=\sum_{n=1}^{\infty}x_n\bar{y}_n$ where $\bar{y}_n$ is the conjugate of $y_n$.   I denote the norm associated with this scalar product by $\| .\|_2$, and c ($l^2, \|.\|_2$) is a Hilbert space. $Z_i$ are $iid$ elements of $l^2$ and $E\left( \| Z_i()\|_2\right)< \infty.$

The application of Assumption 4 leads to $$ f(Z)=\sum_{j=1}^{\infty} Z(j)\pi_j.$$ 

The set of instruments is  infinite and can be written as $Z$. It can be shown that
\begin{equation}
    \frac{f(Z)'f(Z)}{N}=\frac{[\sum_{j=1}^{\infty} Z(j)\pi_j]'[\sum_{j=1}^{\infty} Z(j)\pi_j]}{N}=\pi'\frac{Z'Z}{N}\pi
\end{equation}
with $\pi$ a vector of $G$ sequence in $l^2$.
Thus, \begin{equation}
\pi'\frac{Z'Z}{N}\pi=\frac{\sum_{i=1}^N\pi'Z_i'Z_i\pi}{N}.    
\end{equation}

Note that $\pi_gZ_i'Z_i\pi_g=<Z_i\pi_g, Z_i\pi_g>=\|Z_i\pi_g\|_2$, where $\pi_g$ is an element of $l^2$, $g=1,..,G$. The covariance operator for this sequence is  \begin{equation} \label{operatl}
\mathbf{\Upsilon}:l^{2}\hspace{0.03in}\rightarrow \hspace{0.03in}l^{2}
\end{equation}%
\begin{equation*}
(\mathbf{\Upsilon}\Lambda)_k=\sum_{n=1}^{\infty} E(Z_{i}(k)\overline{Z_{i}(n)})\Lambda(n)
\end{equation*}%

Applying Assumption 5 to $\mathbf{\Upsilon}_l$ means that $\sum_{n=1}^{\infty} E(Z_{in}\overline{Z_{in}}\pi_{gn}^2)<\infty$, and thus $E(\|Z_i\pi_g\|_2)<\infty.$ The application of the weak law of large numbers gives that $\frac{\sum_{i=1}^N\pi'Z_i'Z_i\pi}{N}$ converges in probability $Q$, where $Q$ is a finite positive definite matrix. 

This shows that Assumption 5 implies Assumption 3 ($ii$).

\textbf{($iii$) Continuum of instruments}

The presence of a continuum of moment conditions means that the first-stage representation does not directly have a linear representation. The corresponding representation of Assumption 3 ($ii$)  is that $\frac{f'(Z)f(Z)}{N} \rightarrow Q$.  

$f(Z)$ belongs to the closure of the linear span of  \{$Z(\tau)$ with $\tau \in \mathbb{R}$\} from Assumption 4. Thus, there exists $p(\tau)$ with $\tau \in \mathbb{R}$ such that 
\begin{equation}\label{func_cont}
f(Z_i)= \int Z_i(\tau) p(\tau) d\tau.     
\end{equation} 

Using this general representation, it can be seen that 
\begin{equation}
 \frac{f(Z)'f(Z)}{N} = \int\frac{ p'(\tau_1) Z'(\tau_1)Z(\tau)p(\tau)}{N} d\tau d\tau_1.  
\end{equation}
Assumption 5 claims that $\mathbf{\Upsilon}$ is a compact operator also, I consider $Z_i(\tau) \in L^2(\mu)$ and are $iid$ with $E(\|Z_i(.)\|)<\infty$ and I apply the law of large numbers on the quantify $\frac{ p'(\tau_1) Z'(\tau_1)Z(\tau)p(\tau)}{N}$. Assumption 5 implies that $$ \int\frac{ p'(\tau_1) Z'(\tau_1)Z(\tau)p(\tau)}{N} d\tau d\tau_1 \rightarrow Q.$$
Thus, Assumption 5 implies Assumption 3 (ii). This ends the proof of Proposition 2. \\

The above proposition suggests that a key assumption of the Carrasco asymptotic approximation implies a controversial part of Assumption 3, which is popular in the Bekker asymptotic approximation. The restriction of compactness of the covariance operator ensures some ideal properties of the set of instruments and the first stage equation exist.  

For Assumption 3 to be coherent, additional restrictions should be imposed on the parameters of the first equation.   Note that $\pi$ needs to be estimated, so a researcher should impose reasonable restrictions on the set of instruments, rather than on the coefficient to be estimated.  

Assumption 5 is defined in a specific topological environment (that is, the norm used is clearly defined). This means that in Carrasco's asymptotic approximation, the results obtained depend on the norm used on the space spanned by instruments, which is restrictive. In the estimation, the infinite dimensions of the first stage implies that  problem needs to be regularized to obtain a solution.  The choice of the  regularization parameter is an additional challenge.

To illustrate the use of a regularized estimator, consider the ridge regularization applied to a model with multicollinearity or near-multicollinearity.  I assume that there are many IVs and that the same IVs are used for the 2SLS estimator.

The regularized counterpart of the projection matrix is
$P^{\alpha}=Z([(Z'Z)^2+\alpha I_K]^{-1}[Z'Z])Z'$, with $\alpha>0$.

Given that $Z'Z$ is a positive semi-definite matrix, it can have the spectral decomposition $Z'Z=U'\Lambda U$, where $\Lambda$ is a diagonal matrix of eigenvalues and $U$ is a matrix orthonormal eigenvector; thus, $P^{\alpha}=ZU'([\Lambda^2+\alpha I_K]^{-1}\Lambda)UZ'$.
The regularized 2SLS estimator proposed by Carrasco (2012) is  $$\hat{\delta}_R=\left(X'U'\Lambda_{\alpha}UX\right)^{-1}X'U'\Lambda_{\alpha}Uy$$ with $\Lambda_{\alpha}=[\Lambda^2+\alpha I_K]^{-1}\Lambda$.

The regularized estimator depends on Assumption 5 via $\Lambda_{\alpha}$ and $U$, and the compactness assumption ensures that these quantities exist asymptotically. Under classical assumptions, this is always the case in finite dimensions. However, in infinite dimensions, the choice of the appropriate norm is crucial to ensure compactness. For instance, if $Z'Z$  approximates an infinite sum such that $K>N$, the choice of norm becomes crucial. Moreover, the presence of a regularization parameter in $\hat{\delta}_R$ is  a potential source of bias. However, its choice can be based on data-driven optimal minimization of the approximated mean-squared error (MSE). In \cite{CARRASCO2012}, such procedures achieve smaller MSEs than standard 2SLS estimators. 

While the compactness of the covariance operator enables the development of a coherent many instruments asymptotic framework, it has some limitations, such as the type of instruments that can be used. The following proposition presents the example of an inadmissible IV set.

\begin{prop}
If I consider the set of instrumental variables (IVs) that are orthonormal, Assumption 5 implies that the number of effective IVs is finite.
\end{prop}

\textbf{Proof of Proposition 3}\\
Let us assume that I have an infinite number of instruments. This will correspond to two cases:  $(ii)$ infinite and countable or $(iii)$ continuum of instrumental variables. For each individual, the set of $\{Z_i()\}$ belongs to an Hilbert space. In this case,  $\mathbf{\Upsilon}$ the covariance operators as define in (\ref{operatL}) and (\ref{operatl}) would be the identity operator. It can be verified that the  identity operator is not compact.  Leading to a contradiction. This ends the proof of Proposition 3. \\

The compactness assumption may also fail to hold if all instruments are orthogonal with $L^2(\mu)$ or $l^2$ norm. However, in the case of finite number of instrumental variables, the covariance operator is compact as a finite-dimensional matrix. 

The result in Proposition 3 suggests that regularization methods that rely on the Assumption 5 for asymptotic approximation regularities should in practice perform better if the IVs are correlated.

\section{Conclusion}

Many IVs are regularly used in empirical studies (see \cite{hansen2008estimation} for some examples). In finite sample estimation, the popular 2SLS estimator has a bias that increases with the number of instruments.  This note shows that a crucial assumption of Bekker's asymptotic approximation  may have strong limiting consequences. Indeed,  in the Bekker's asymptotic approach, the number of effective IVs is should be finite.  This restricts its application of this asymptotic framework to the case of many weak instruments  or sparse set of instrumental variables.  This  may explain the poor behavior of the many IVs asymptotic approaches noted by \cite{harding2016finite}. 

The investigation of Carrasco's asymptotic framework shows the importance of the covariance operator compactness assumption. The covariance compactness assumption implies the controversial crucial part of Bekker's asymptotic framework. Moreover, this assumption restricts the IV set, as it depends on a specific topology imposed on the space spanned by instruments. The norm used should be clearly specified, because in infinite dimensions, all norms are not equivalent.  The compactness assumption does not support orthonormal IV sets unless the number of IVs is finite.  Future research should consider the extension of finite sample approximation, following \cite{harding2016finite}, to regularized k-class estimators.

\bibliography{ManyInstbib}
\bibliographystyle{econometrica}

\end{document}